# Multi-dimensional wave steering with higher-order topological phononic crystal

Changqing Xu[1]†, Zeguo Chen[2]†, Guanqing Zhang[2]†, Guancong Ma[2*], Ying Wu[1*]

[1]Division of Computer, Electrical and Mathematical Science and Engineering, King Abdullah University of Science and Technology (KAUST), Thuwal 23955-6900, Saudi Arabia

[2]Department of Physics, Hong Kong Baptist University, Kowloon Tong, Hong Kong

Emails: Ying.Wu@kaust.edu.sa, phgcma@hkbu.edu.hk

† These authors contributed equally.

**Abstract:**

The recent discovery and realizations of higher-order topological insulators enrich the fundamental studies on topological phases. Here, we report three-dimensional (3D) wave-steering capabilities enabled by topological boundary states at three different orders in a 3D phononic crystal with nontrivial bulk topology originated from the synergy of mirror symmetry of the unit cell and a non-symmorphic glide symmetry of the lattice. The multitude of topological states brings diverse possibility of wave manipulations. Through judicious engineering of the boundary modes, we experimentally demonstrate two functionalities at different dimensions: 2D negative refraction of sound wave enabled by a first-order topological surface state with negative dispersion, and a 3D acoustic interferometer leveraging on second-order topological hinge states. Our work showcases that topological modes at different orders promise diverse wave steering applications across different dimensions.

Since its incipience in the 1980s, the study of topological notion in physics has attracted tremendous attention across diverse disciplines[1-3]. We have witnessed the emergence of a kaleidoscope of topological phases, such as quantum Hall system[1-3], Chern insulators[4-6], and the relatively recent discovery of higher-order topological phases[7-14]. As a hallmark, the nontrivial topology of the system can lead to the existence of certain types of excitations that are localized at the boundaries of the system. Some of these topologically protected states have fascinating properties, such as being immune to backscattering[15-25].

While many topological phases originate from the electronic systems[1-3], they quickly gained the attention from other realms, spanning from optics, photonics[26,27], electromagnetism[4-6,15-17], to acoustics and phononics[18-25]. Due to the versatility offered by these classical systems, they rapidly become platforms to realize novel topological phases and to investigate the physics therein. However, relatively few efforts were devoted to the exploration of novel topological states for wave manipulation applications. This work is devoted to applying topological states to achieve novel wave steering at different dimensions. It is based on a simple realization of a 3D higher-order topological phononic crystal (PC) possessing a large bandgap that can be characterized by the nontrivial quantized bulk polarization. Differs from the acoustic analog of Su-Schrieffer-Heeger (SSH) model[10] or Kagome model[8], our PC maintains nontrivial topology after a reversion of the center and corner of the unit cell, which means our PC does not have a topologically trivial counterpart. As a result of a ternary layer of topological protection, a hierarchy of 2D topological surface states (TSSs), 1D topological hinge states (THSs), and 0D topological corner states (TCSs) are observed. We then present the dispersion of the TSSs and show that it can lead to negative refraction at PC-air interfaces. In addition, we exploit the THSs as tailorable transport channels to realize a 3D acoustic interferometer. Our work showcases that topological states can be tailored for diverse and versatile wave steering applications across multiple dimensions.

**Results**

**Phononic crystal with higher-order topology.** Our PC comprises a cubic array of orthogonally aligned aluminum rods (treated as sound-hard objects) along the *x*-, *y*-, *z*-directions, respectively, in an air background. All rods have a square cross-section with a side length $L = 1.8$ cm, and their axes are separated by $a/2$, where $a = 4$ cm is the lattice constant. The PC belongs to a non-symmorphic space group no. 223[28] and has a glide symmetry $G_{\mu\nu} = \left\{M_{-\mu\nu} | \frac{a}{2}\hat{\xi}\right\}: (\mu, \nu, \xi) \to \left(-\nu, -\mu, \xi + \frac{a}{2}\right)$, where the subscripts represent spatial coordinates. It transforms a rod at $\mu = (x, y, z)$ to $\nu = (y, z, x)$ through a reflection over $(-\mu, \nu)$ plane followed by a translation of $a/2$ along $\xi = (z, x, y)$[29-31]. A selection of unit cell labelled as "uc1" is made such that mirror symmetry $m_\mu$ with respect to all three bisecting planes. The PC's glide symmetry implies that shifting the unit cell along one of the diagonal directions by a distance $(\Delta x, \Delta y, \Delta z) = \left(\frac{a}{2}, \frac{a}{2}, \frac{a}{2}\right)$ produces a different choice of unit cell also possessing the same mirror symmetry $m_\mu$. This unit cell's center coincides with the corner of the original unit cell. For presentation purpose, we shift the uc1 by $(\Delta x, \Delta y, \Delta z) = \left(\frac{3a}{2}, \frac{a}{2}, \frac{a}{2}\right)$, and label the new unit cell as "uc2" in Fig. 1a.

Figure 1b shows the band structure along high symmetry lines of the first Brillouin zone. The combination of lattice symmetry and geometry parameters results in a large complete bandgap from 5.0 kHz to 6.8 kHz. Protected by the glide symmetry, the two bands below the bandgap are doubly degenerate at the Brillouin zone boundaries[30,32]. Meanwhile, the unit cell's mirror symmetry indicates the possibility of quantized band topology[10]. Here, we characterize the band topology using the Wannier bands[33], which consists of Wannier centers $\boldsymbol{v}$ as functions of wave vector $\boldsymbol{k}$. The Wannier centers are the phases of the eigenvalues of the Wilson loop operator in the Brillouin zone (see details in the method). Our computation yields two gapped and quantized Wannier centers $\boldsymbol{v^1} = (0,0,0)$ and $\boldsymbol{v^2} = \left(\frac{1}{2}, \frac{1}{2}, \frac{1}{2}\right)$ for these two bands (Fig. 1c), indicating non-zero quantized bulk dipole moment $\boldsymbol{P} = \left(\frac{1}{2}, \frac{1}{2}, \frac{1}{2}\right)$. Consistent results can be obtained by analyzing the parities of Bloch wavefunctions at the high-symmetry points in the Brillouin zone[10,34]. The Block wavefunctions at the high-symmetry points labeled in Fig. 1b are plotted in Figs. 1d-k. It is seen that, at the Brillouin zone center, the wavefunctions have identical parity (Figs. 1d, e). In contrast, since the glide symmetry mandates the two bands to be degenerate at the Brillouin zone boundary, the corresponding wavefunctions must have opposite parities. Interestingly, as mentioned before, the glide symmetry also implies that the second choice of the unit cell (uc2), which also preserves mirror symmetry $m_\mu$, gives identical bulk dipole moment $\boldsymbol{P} = \left(\frac{1}{2}, \frac{1}{2}, \frac{1}{2}\right)$. In other words, the nontrivial bulk polarization of the first gap exists for both choices of the unit cell. Consequently, under the premise of $m_\mu$, the glide symmetry of our PC leads to the robustness of bulk polarization to different selections of unit cell. This characteristic distinguishes our PC from certain higher-order topological insulators, such as the 'breathing' Kagome lattice[8] and the 3D SSH model[10], in which the bulk topology depends on the selection of unit cell.

**Negative refraction at PC-air interfaces by TSS.** We consider a lattice truncation, which gives rise to topological boundary modes[8,35,36]. Shown in Fig. 2a is the supercell, consisting of 7 unit cells in the $z$-direction, and periodic in $x$- and $y$- directions. The top unit cell is truncated by a plane at a distance $dh \in [0, 4.0]$ cm away from the top, as illustrated in Fig. 2a. Two sound-hard boundaries cover the top and bottom ends of the supercell. Figure 2b shows the dispersion of the TSS along high symmetry lines of the surface Brillouin zone at $dh = 2.9$ cm. Note that the group velocity of the first TSS is negative, suggesting it can be leveraged for negative refraction.

TSS-enabled negative refraction is demonstrated with the setup shown in Fig. 2c. The PC sample consisting 20×6 identical supercell with fixed truncation position ($dh = 2.9$ cm) is bounded in a cuboid aluminum box. Two planar waveguides of 5 mm thickness are connected to the top surface via two thin slits. The top plate of the waveguides and the PC is removed to show the microphone and the PC. A loudspeaker emits a near-Gaussian beam at a frequency of 6.06 kHz and at an incident angle of 15°. The beam propagates inside the waveguide and incident to the top surface of the PC to excite the TSS. We use a microphone to raster-scan the acoustic field inside both the waveguides on the incident and transmitted sides. The scanning regions are marked by the dashed black boxes. In Fig. 2d, the incident field and transmitted field are plotted, normalized by a rescaling to their maximum, respectively. As expected, the measured acoustic pressure field shows that the output beam is right-shifted, verifying the negative refraction.

**Observation of higher-order topological states.** We consider a PC cube consisting 5×5×5 unit cells (shown in Fig. 3a) enclosed by sound-hard boundaries at all sides. The truncations are chosen as $dh = 1.0$ cm on the three adjacent boundary surfaces (visible in Fig. 3a) and $dh = 3.0$ cm on the rest three boundary surfaces (hidden in Fig. 3a). Such truncation scheme results in two different types of hinges, as respectively indicated by black and cyan dashed lines in Fig. 3a. We calculated the eigenspectrum of this PC cube and plot eigenfrequencies versus the solution number in Fig. 3b, in which first-order TSSs, second-order THSs, and third-order TCSs are identified in the bulk bandgap. The in-gap THSs only exist at the black hinges. Likewise, the in-gap TCSs exist at the crossing point of three cyan hinges. The THSs cluster in a frequency range of 5.0 ~ 5.8 kHz, whereas TCSs are found at 6.69 kHz. The acoustic pressure field distributions at 5.14 kHz and 6.69 kHz are plotted in Figs. 3c and 3d, manifesting the typical THSs' and TCSs' features, respectively.

To verify the simulated results, we fabricated this 5×5×5 PC cube enclosed in aluminum plates, whose picture is shown in Fig. 3e without the top aluminum plate for viewing purposes. Arrays of holes are drilled on the aluminum plates. The holes are blocked when not in use. Sound sources can be placed at different positions and excite the PC cube through the holes. First, we insert a microphone well inside the PC through the holes to measure the bulk response. The normalized pressure field is plotted in the black curve in Fig. 3f, exhibiting a large bandgap in the frequency range 4.9 – 7.0 kHz as predicted by the simulation. Then the hinge response is also observed, which exhibits a dominant peak near 5.14 kHz followed by a plateau extending to about 6.8 kHz, as shown in the green curve in Fig. 3f, implying the existence of localized states on the surfaces or at the hinge. The response at the corner is plotted in Fig. 3f in red curve. Only one dominant peak at 6.72 kHz is observed, agreeing with the numerical simulation. To further confirm the existence of THSs and TCSs, we mapped out the acoustic field distribution on the surfaces, hinges, and corners. The pressure field maps obtained at 5.20 kHz and 6.72 kHz are shown in Figs. 3g and 3h, and clearly demonstrate the localization of the sound wave on the hinges and at the corner, respectively, affirming the coexistence of second-order and third-order topological states in our PC. The stars in Figs. 3g and 3h represent the position of sound sources.

**Wave transport leveraging THSs**. Higher-order topological states give rise to new possibilities for wave manipulation. For example, THSs naturally offer tailorable wave transport channels that can be tailored to versatile shapes. As a proof-of-principle demonstration of potential practical applications of the higher-order topological states, an interferometer based on the PC cube are fabricated and characterized. As illustrated in Fig. 4a, three rectangular waveguides, labeled port 1, 2, and 3, are connected to three corners of the aforementioned 5×5×5 PC cube. The location of these waveguides ensures three-fold rotational symmetry about the diagonal axis of the PC cube. The cross-section of each waveguide is set as 1.6×0.1 cm. Figure 4b shows the simulated acoustic field distribution when a wave with frequency 5.16 kHz incidents from port 1. The THSs are excited and indeed functions as waveguiding channels, which can be exploited to construct an interferometer. Figure 4c plots the result of two in-phase waves incident to the cube from ports 1 and 2. Owing to the constructive interference, the amplitude of the outgoing sound at port 3 is doubled. In contrast, when two out-of-phase waves incident onto the same ports, destructive interference occurs, and consequently, the suppression of the outgoing sound at port 3 is observed, as shown in Fig. 4d. Figure 4e gives a picture of the experimental setup of such a THS interferometer. Two speakers are used as sound sources, placed at the end of two aluminum waveguides with a rectangular cross-section. The other ends of the waveguides are connected to the two corners of the PC cube. We measure the sound amplitude in the output waveguide at the upper right corner. The transmittance for 5.0 – 5.5 kHz in Fig. 4f is peaked at 5.27 kHz when the two speakers are in-phase, while it is minimized at the same frequency when the two speakers are out-of-phase. These results are strong evidence of the constructive and destructive interferences attributed to two separate but equivalent hinge paths.

**Discussion and Conclusions**

We present a simple design of an acoustic 3D high-order TI that can simultaneously support topological states at three different dimensional hierarchies. The negative dispersion of the TSS makes negative refraction easily attainable. Note that our negative refraction occurs at the PC-air interfaces instead of between two different PC boundaries, making it feasible for wave-steering applications. The presence of higher-order topological states, in particular, the THSs, brings even more intriguing effects. First, THSs along different hinges can relay the transport of sound waves, guiding the propagation to bend around corners towards different directions. As a result, the inputs and output are not on the same plane. Similar effects can only be attained by using a 3D double-zero-index medium previously[31], which relies on the stringent tuning of system parameters. In comparison, the topological protection of THSs endows additional robustness. Second, it is easy to see the design of the THS interferometer affords great versatility. For example, by choosing the position of input and output ports, the interferometer can easily guide waves towards different directions in the 3D space (as shown in Fig. 4b). Such functionality is unobtainable for any 2D TIs.

In summary, our work presents convincing cases that higher-order topological wave crystals can benefit wave-steering applications. As topological notions can be universally applied to other realms of physics, such as photonics and electromagnetism, we believe our work is an important step for a broad area of next-generation technology and devices.

**Methods**

**Simulations.** The band structures, Bloch wavefunctions, and acoustic field distributions were calculated using the acoustic module in COMSOL MULTIPHYSICS. The Wannier centers are obtained by computing the Wilson loop operator defined as:

$$\mathcal{W}_\mu(\mathbf{k}) = -\frac{i}{2\pi} \int_0^{2\pi/a} u^{m,\dagger}(\vec{k}) \partial k_\mu u^n(\vec{k}) dk_\mu, \qquad (1)$$

where $m$ and $n$ are the band indices, $\mathbf{k}$ defines the starting point of the loop, $\mu = x, y, z$ indicates the loop direction, $u^n(\vec{k})$ is the periodic part of the Bloch wavefunction. The integration is carried out by discretizing the Brillouin Zone into 20 segments in each direction of $\vec{k}$. We then diagonalize the Wilson loop operator $\mathcal{W}_\mu(\mathbf{k})$ as $\mathcal{W}_\mu(\mathbf{k})|v_k^m\rangle = e^{i2\pi v_k^m}|v_k^m\rangle$. The phase $v_k^m$, which depends on the band index and the Wilson loop, is the Wannier center of the Bloch wavefunctions. As we choose the loop direction along the *x-, y-, z*-directions, respectively, the numeric results show gapped Wannier bands and $v_k^m$ is 0 or 0.5, depending on the Wannier band index.

**Experiments.** All PC samples and their claddings were machined from aluminum alloy and then were assembled in the lab. The frequency responses were measured by frequency scans. A waveform generator (Keysight 33500B) was used to generate a monochromatic signal to drive a loudspeaker through an audio power amplifier. The signals were detected by a 1/4-inch microphone (PCB Piezotronics Model-378C10) and were then recorded by a digital oscilloscope (Keysight DSO2024A). For the negative refraction experiments, the raster-scans of acoustic fields were performed using a house-built motorized translation stage. For the measurement of THSs and TCS, manual scans were performed by inserting the microphone to all small ports on the claddings (Fig. 3e).

**Acknowledgments**

Y. W. was supported by the King Abdullah University of Science and Technology (KAUST) Office of Sponsored Research (OSR) under Award No. OSR-2016-CRG5-2950 and KAUST Baseline Research Fund BAS/1/1626-01-01. G. M. was supported by the Hong Kong Research Grants Council (GRF 12302420, 12300419, ECS 22302718, CRF C6013-18G National Science Foundation of China Excellent Young Scientist Scheme (Hong Kong & Macao) (#11922416) and Youth Program (#11802256), and Hong Kong Baptist University (RC-SGT2/18-19/SCI/006).

**Author contributions**

C. X. designed the phononic crystal. C. X. and Z.-G. C. performed the numerical simulations and designed the experiment. Z.-G. C. and G. Z. set up the experiment and carried out the measurements. C. X., Z.-G. C., G. M. and Y. W. wrote the manuscript with inputs from all authors. The project was supervised by G. M. and Y. W.

**Competing interests:** The authors declare no competing interests.

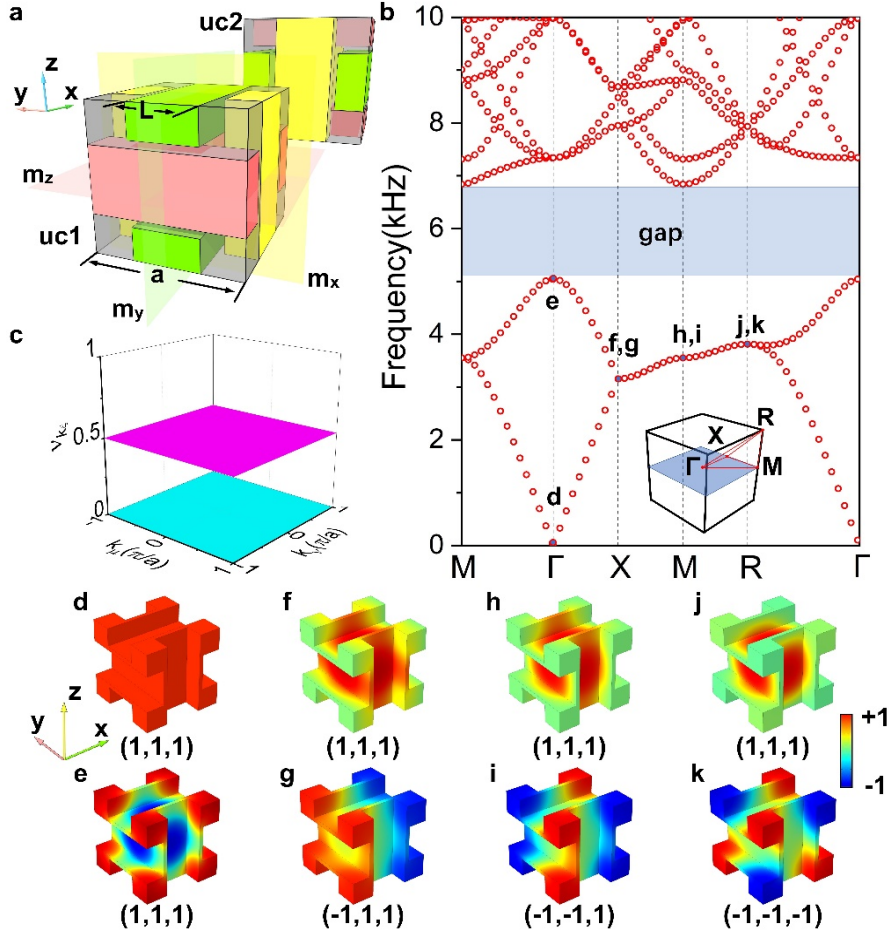

**Fig. 1. The design and characteristics of the PC. a** The unit cell with a lattice constant $a = 4$ cm. The host medium (gray) is air. Uc1 and uc2 are different choices of the unit cell, possess the same mirror symmetries. **b** Band structure of the PC with $L = 1.8$ cm. **c** The bulk polarization of the first two bands below the bandgap. **d-k** Acoustic pressure field distributions of the eigenstates at the **d, e** Γ point, **f, g** X point, **h, i** M point and **j, k** R point of the first and the second bands. The eigenvalues of eigenstates under mirror operator $m_\mu$ are shown below.

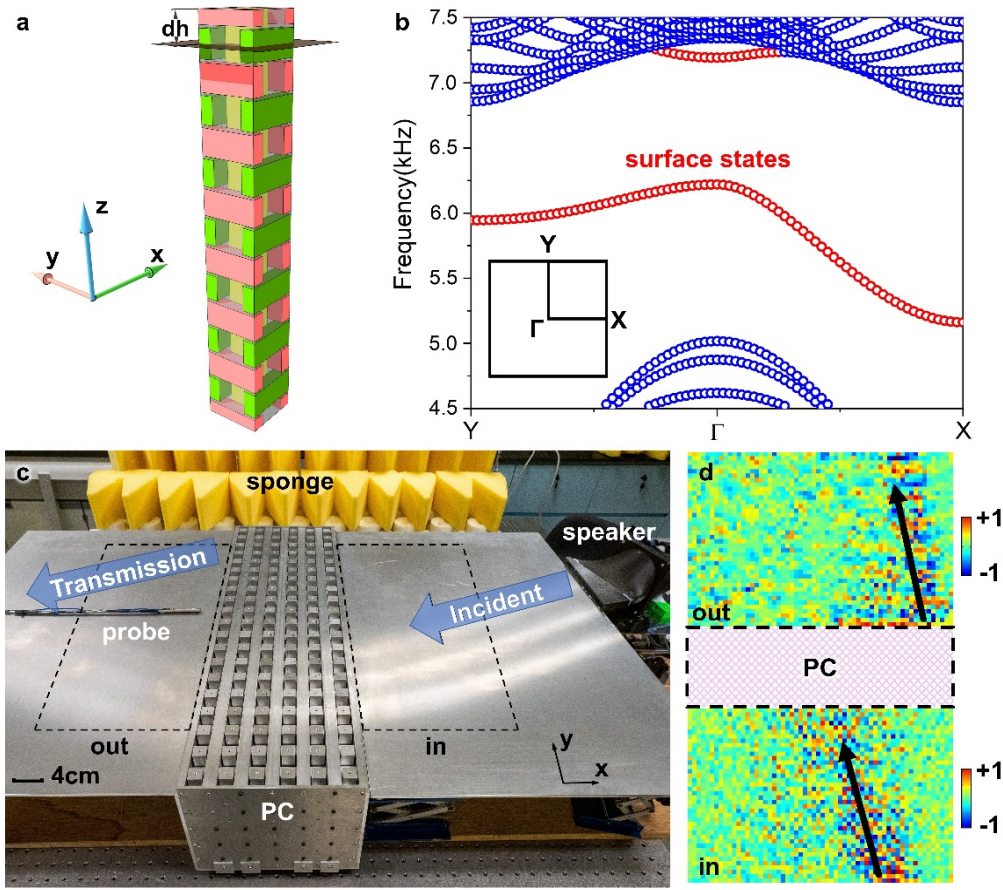

**Fig. 2. The negative refraction based on a topological surface state. a** The top of the supercell is truncated by a sound-hard boundary (the black plane), while the bottom is bounded by another. **b** Calculated band structures in the surface Brillouin zone (inset) at $dh = 2.9$ cm. **c** Experimental setup to realize negative refraction at the surface of PC with $dh = 2.9$ cm. Acoustic wave incident with frequency 6.06 kHz and incident angle 15°. **d** Experimental results of acoustic pressure field in the dashed boxes in **c**.

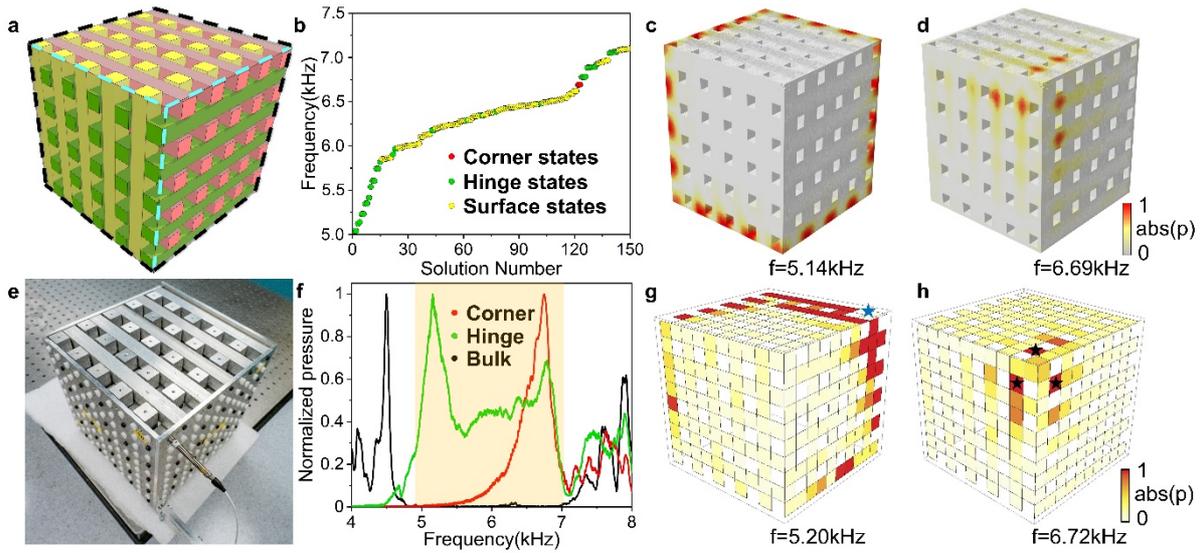

**Fig. 3. Higher-order topological states in a PC cube. a** The PC cube is composed by 5×5×5 units, truncated by sound-hard boundaries in all directions. **b** The eigenspectrum of the PC cube, exhibits a hierarchy of surface states (yellow), hinge states (blue) and corner states (red). The acoustic pressure field of **c** hinge state and **d** corner state, are plotted. **e** Experimental setup. The aluminum wall at the top is removed to show the PC. **f** Measured normalized pressure when the probe is placed at the bulk (black), hinge (green) and corner (red). The shadow indicates the frequency range of bandgap. **g** Acoustic pressure field of a hinge state at frequency 5.20 kHz. **h** Acoustic pressure field of a corner state at frequency 6.72 kHz. The stars in **g** and **h** indicate the position of the source.

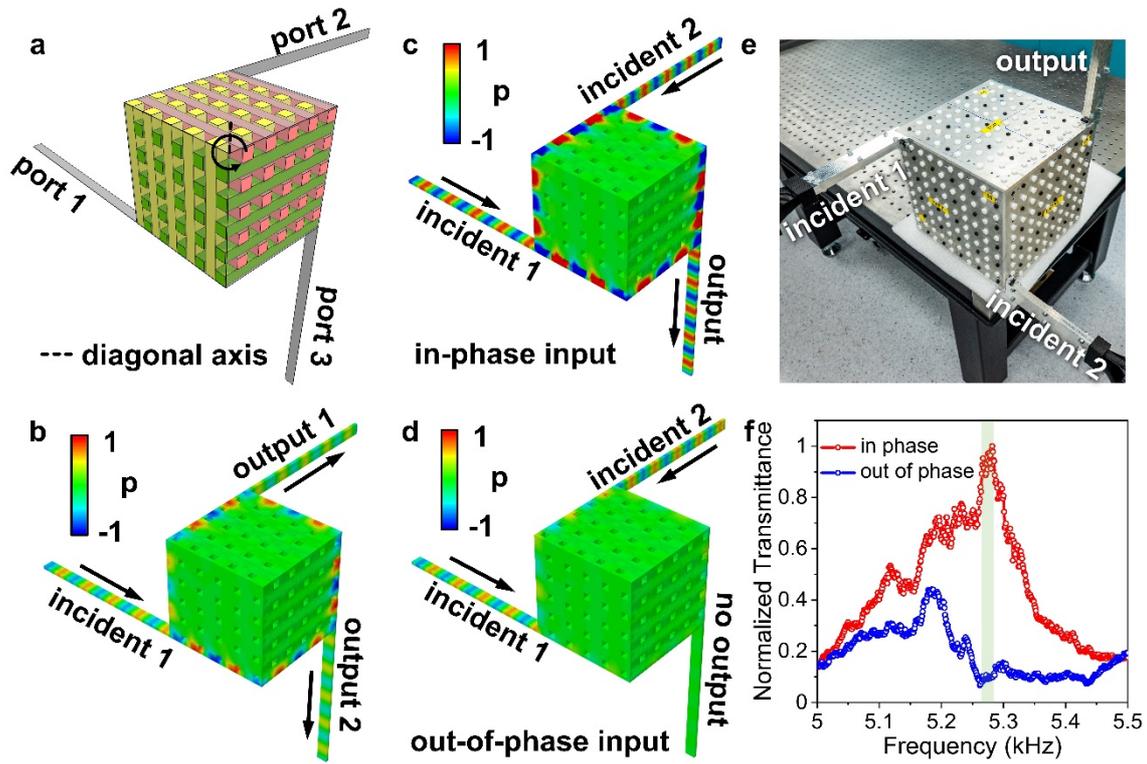

**Fig. 4. A 3D interferometer based on topological hinge states. a** Three ports are connected to the hinges of the PC cube in Fig. 3. **b** The hinge states are excited when wave incident in port 1. **c** The constructive interference when in-phase signals incident in port 1 and port 2. **d** The destructive interference when out-of-phase signals incident in port 1 and port 2. **e** Experimental setup. The incident wave comes from two speakers with configurable phases. The sound signals are detected via the waveguide at the upper right corner. **f** Normalized transmittance from the probe with in-phase (red) and out-of-phase (blue) inputs.